# A Backscattering Model Incorporating the Effective Carrier Temperature in Nano-MOSFET

Gino Giusi [(a)], Giuseppe Iannaccone [(b)], Felice Crupi [(a)], Umberto Ravaioli [(c)]

[(a)] Dipartimento di Elettronica, Informatica e Sistemistica, University of Calabria, 87036 Arcavacata di Rende, Italy (e-mail: ggiusi@deis.unical.it, crupi@unical.it)

[(b)] Dipartimento di Ingegneria dell'Informazione, Elettronica, Informatica, Telecomunicazioni, Università di Pisa, 56126 Pisa, Italy (e-mail: giuseppe.iannaccone@unipi.it)

[(c)] University of Illinois at Urbana-Champaign. Urbana, IL 61801 USA (ravaioli@uiuc.edu)

## Abstract

In this work we propose a channel backscattering model in which increased carrier temperature at the top of the potential energy barrier in the channel is taken into account. This model represents an extension of a previous model by the same authors which highlighted the importance of considering the partially ballistic transport between the source contact and the top of the potential energy barrier in the channel. The increase of carrier temperature is precisely due to energy dissipation between the source contact and the top of the barrier caused by the high saturation current. To support our discussion, accurate 2D full band Monte Carlo device simulations with quantum correction have been performed in double gate nMOSFETs for different geometries (gate length down to 10 nm), biases and lattice temperatures. Including the effective carrier temperature is especially important to properly treat the high inversion regime, where previous backscattering models usually fail.

# I. Introduction

Modeling ballistic and partially ballistic charge transport in nanoscale transistors is of fundamental importance given the relentless scaling down of device size and has attracted significant research interest in recent years [1-14]. Modeling of quasi-ballistic transport is commonly faced with the Lundstrom model (LM) [2] which is based on the simplified theory of the ballistic transistor developed by Natori [1]. In the LM charge transport in the channel is regulated by the injection of thermal carriers from the top of the barrier (called the "virtual source" or VS) into the channel. This picture has provided a radically new way to look at the problem, because it moved attention from the drain to the source. However the model attracted some criticisms from the literature regarding the assumption of elastic transport as well as the specific expression for the backscattering coefficient [3, 4, 15]. Recently we proposed a model for charge transport in the channel which retains the useful features of the approaches in [1] and [2] but avoids the concept of virtual source, and includes a less idealized picture of transport between the source contact and the potential energy peak in the channel (the former VS) [14]. In this way, for the partially ballistic regime, charge distribution at the potential energy peak is not represented by the superposition of hemi-maxwellians. In this work we show that, in the saturation regime, carriers at the top of the barrier are not in equilibrium with the lattice and have a higher effective temperature. The inclusion of the effective carrier temperature allows us to extend the range of applicability of the previously proposed model [14], in terms of device geometry, bias dependence and temperature.

# II. Device Structure and Simulation Method

The simulated reference device is a double-gate (DG) n-type metal-oxide-semiconductor field effect transistor (nMOSFET) with a scaled gate length, an ultra-thin un-doped silicon body, an oxide ($SiO_2$) thickness of 1.5 nm and long (35 nm) source and drain heavy doped ($10^{20}$ $cm^{-3}$) n-type regions (Fig. 1). The use of these long regions, which are not part of the effective device and require an additional computational cost, is necessary to avoid artifacts in the carrier distributions injected by source and drain contacts. Let us stress the fact that the source and drain contacts - which are assumed here as ideal reservoirs - are external to the simulation domain. Therefore the source and drain (n+) regions in the simulation domain are not part of the reservoirs, and experience non-equilibrium transport and dissipation. Gate length is varied from 10 to 50 nm and silicon thickness is varied from 1 to 10 nm. The small silicon body thickness is selected to approach the one-dimensional (1D) electrostatics which is a basic assumption for the transport model used in this work. The simulation tool is a two-dimensional (2D) full band self-consistent Monte Carlo (MC) device simulator for Silicon. It includes several scattering mechanisms such as phonons, ionized impurities, and surface roughness [16]. The simulation is semi-classical and carrier energy is distributed according to a three-dimensional (3D) density of states ($g_{3D}$). Quantum confinement effects are taken into account through an efficient correction to the electrostatic potential [17].

# III. Proposed Model for Channel Backscattering

In Ref. [14] we avoided the concept of VS as defined by the LM [2]: we assume that in saturation carriers are injected into the channel from the source reservoir with an equilibrium distribution and then we include in the model the effect of collisions occurring between the source contact and the "top of the barrier" or $x_{max}$ (which is called VS in the LM). In saturation

$$I_D \approx \frac{1-r}{1+r} I_{S,BL}^{+} \tag{1}$$

$$Q \approx \left(1 + rk_v\right) n^+_{S,BL} \qquad (2)$$

where $I_D$ is the drain saturation current, $Q$ is the charge density at $x_{max}$, $k_v$ is the ratio of the average velocity of positive directed carriers $v^+$ to the average velocity of negative directed carriers $v^-$ at $x_{max}$, $r=I^-/I^+$ is the backscattering ratio (BR) as defined in the LM [2], $I^+$ $(I^-)$ is the positive (negative) directed current at $x_{max}$ and $\left(I^+_{S,BL}, n^+_{S,BL}\right)$ are the positive directed moments at $x_{max}$ which are functions of the carrier temperature $T_C$ and of the normalized electrostatic potential $\eta = \left(E_{FS} - E_C\right)/kT_C$ where $E_{FS}$ is the source reservoir Fermi level, $E_C$ the conduction band energy at $x_{max}$ and $k$ is the Boltzmann's constant. In the Natori-Lundstrom picture, carriers at $x_{max}$ are assumed in equilibrium with the lattice so that the carrier temperature is assumed equal to the lattice temperature ($T_L$). The assumption of carrier temperature $T_C = T_L$ has also used also in our previous model [14]. It is strictly correct at equilibrium when $V_{DS}$=0 and $I_D$=0, but when a strong current flows in the channel, energy dissipation between the source contact and $x_{max}$ increases carrier kinetic energy, so that the effective temperature of carriers at $x_{max}$ is higher than $T_L$. Assuming that carriers are distributed following a Fermi-Dirac function at $x_{max}$, the effective carrier temperature $T_C$ is linked to the average carrier energy $E$ and to the normalized electrostatic potential $\eta$ evaluated all at $x_{max}$, by

$$E = g\,\frac{kT_C}{2}\,\frac{\Im_{g/2}\left(T_C,\eta\right)}{\Im_{g/2-1}\left(T_C,\eta\right)} \qquad (3)$$

where $g$ is the system dimensionality and $\Im_j$ the Fermi-Dirac integral of order $j$. The ratio of Fermi-Dirac integrals in Eq. (3) accounts for degeneracy. The model proposed in this work can be represented by the system of Eqs. 1-3, three equations in the three unknowns $r$, $T_C$, $\eta$. The values for $I_D$, $Q$ and $E$ at $x_{max}$ as well as $k_v$ are extracted from the Monte Carlo simulation. In [3] and [14], $k_v$ was fixed to a value 1.35. To be consistent with our MC device simulation tool, and deviating from Ref. [14] where 2D ballistic equations have been used, here we adopt 3D equations for the ballistic moments ($g$=3)

$$J^{+}_{S,BL} = q\left(\frac{2\pi m_{DOS}kT_C}{h^2}\right)^{3/2}\sqrt{\frac{2kT_C}{\pi m_C}}\Im_1(\eta) \tag{4}$$

$$n^{+}_{S,BL} = \left(\frac{2\pi m_{DOS}kT_C}{h^2}\right)^{3/2}\Im_{1/2}(\eta) \tag{5}$$

where $J^{+}_{S,BL}$ is the ballistic source-injected current density (A/cm$^2$) and $h$ is the Planck's constant. The density of states effective mass $m_{DOS}$ and the conduction effective mass $m_C$ are 1.08 $m_0$ and 0.26 $m_0$, respectively, where $m_0$ is the free electron mass. For experimental extraction the method proposed in [13] with the correct model presented here may be used, in conjuction with an approximated method to extract carrier temperature from experiments.

# IV. Monte Carlo Device Simulation Results

The result of MC device simulation in a device structure with $L$=20 nm, $t_{si}$=7.5 nm, $t_{ox}$=1.5 nm at the lattice temperature $T_L$=300 K for different gate voltages and $V_{DS}$=1 V is shown in Fig. 2. In this figure the backscattering ratio (BR) $I^-/I^+$ extracted from the MC has been compared with the values calculated with the proposed model and with the LM. For both models the cases $T$=$T_L$ (as reported in [14]) and $T$=$T_C$ (as proposed in this work) have been considered (in the case of the LM the correct $k_v$ extracted from MC has been considered in order to exclude this source of error). In the same figure the effective carrier temperature at $x_{max}$, $T_C$, calculated by the proposed method, has been plotted as a function of gate voltage. For low gate voltages $T_C \approx T_L$ and the model reported in [14] works well. However, when the overdrive increases, carriers at $x_{max}$ are not in equilibrium with the lattice and $T_C$ significantly departs from $T_L$. Correspondingly, the model reported in [14] fails while the model proposed in this work, where the effect of the effective temperature has been included, continues to work quite well. It is worth noting that the LM is not able to accurately reproduce the BR also in the case in which the effect of $T_C$ is included, thus confirming that neglecting the scattering between the source and $x_{max}$ is a major source of error in the backscattering calculation, as discussed in [14]. Figure 3 shows the BR calculated for different geometries, biases and lattice temperatures. When not changed, $t_{si}$ = 1.5 nm, $L$ = 20 nm, $V_{GS}$ = 1.4 V and $T_L$ = 300 K. The silicon thickness is chosen to be very small in order to emphasize the disequilibrium at $x_{max}$. Reducing $t_{si}$ is equivalent to increasing the gate voltage because both degenerate conditions and current density increase. As a result, also the carrier temperature increases. It is evident that the predictions of the proposed model are in close agreement with the values obtained by MC simulation in all cases, while the other models show relevant error in backscattering calculation. Fig. 4 shows the carrier energy distribution at $x_{max}$ as extracted from the MC and the one calculated using a Fermi-Dirac function with $T_C$ and $\eta$ extracted by our model. The good agreement confirms our hypothesis that, in saturation, carriers are heated at $x_{max}$ with a temperature remarkably higher with respect to the lattice temperature.

This study demonstrates the necessity of including the effective carrier temperature at the top of the barrier and dissipation from the source contact to the top of the barrier to accurately model channel backscattering in the high-inversion regime where previous backscattering models usually fail.

# Figures

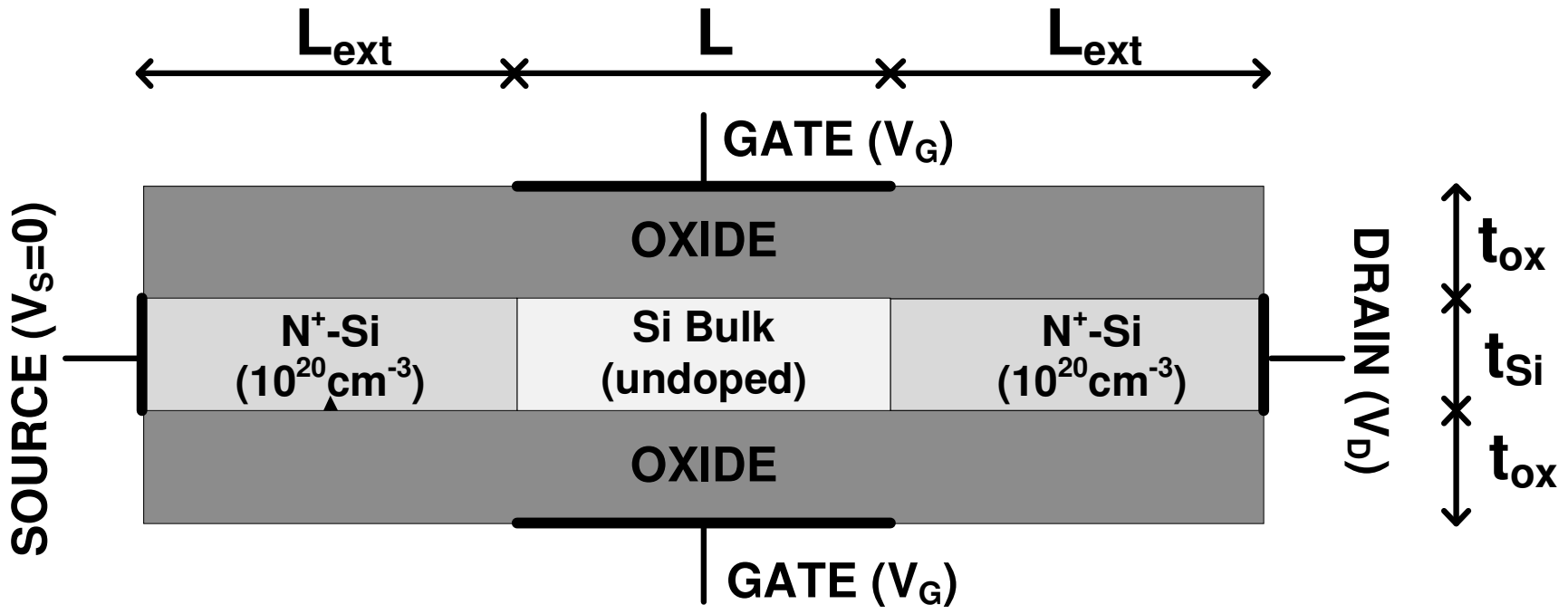

Figure 1. The simulated structure is a double gate nMOSFET with thin undoped silicon body, oxide thickness $t_{ox}$=1.5 nm and long source/drain extensions ($L_{ext}$=35nm).

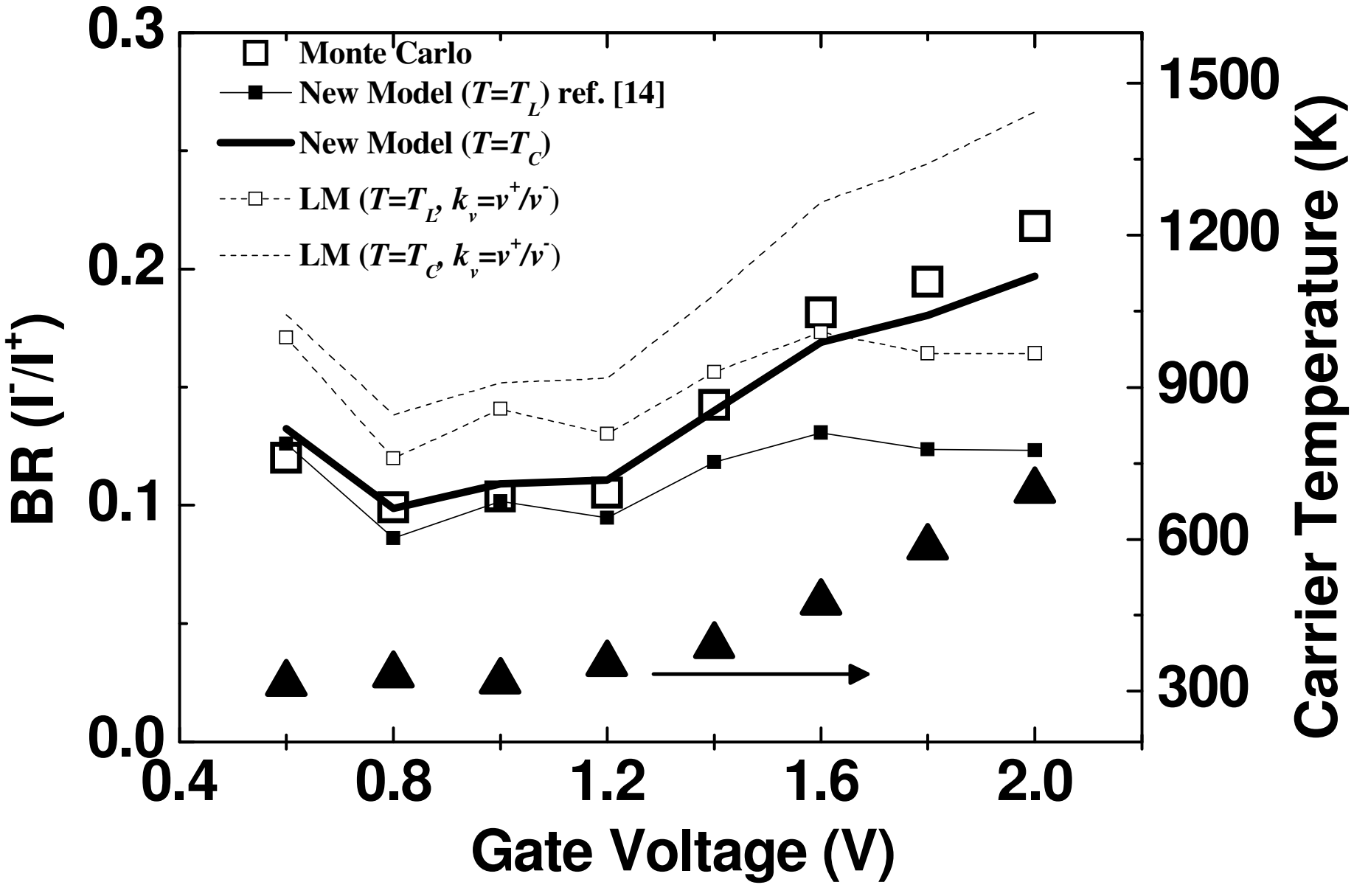

Figure 2. The BR (left axis) and the effective carrier temperature $T_C$ (right axis) in a device structure with $L$=20 nm, $t_{si}$=7.5 nm, $t_{ox}$=1.5 nm at the lattice temperature $T_L$=300 K for different gate voltages and $V_{DS}$=1 V.

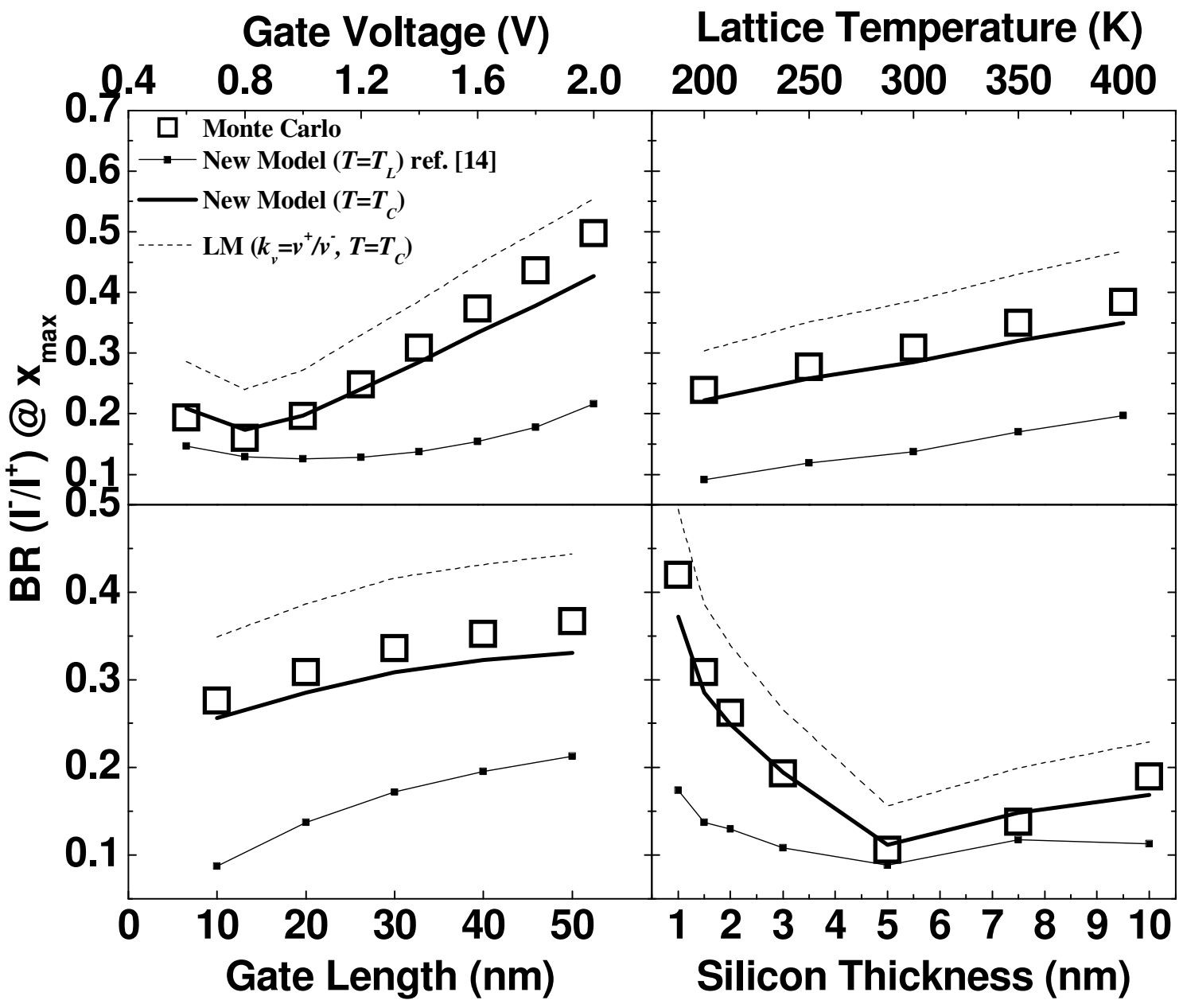

Figure 3. The BR as a function of device geometry, bias and lattice temperature with $V_{DS}$=1V (when not changed, $L$=20 nm, $t_{si}$=1.5 nm, $V_{GS}$=1.4V, $T_L$=300K). The proposed model allows us to obtain a more accurate reproduction of the MC simulation results with respect to previous models for all cases.

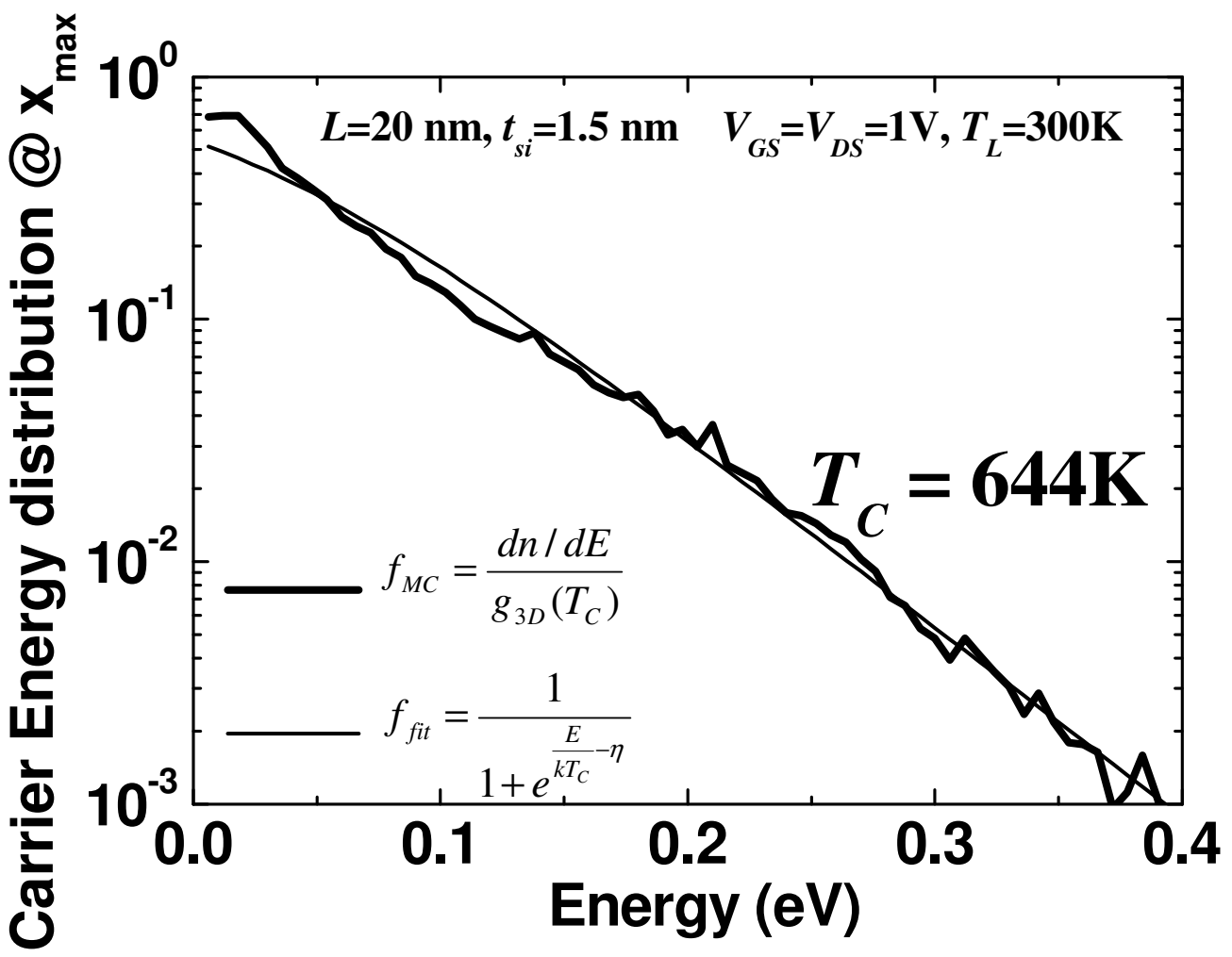

Figure 4. The carrier energy distribution at $x_{max}$ extracted from the MC ($f_{MC}$) and the one calculated with a Fermi-Dirac function ($f_{fit}$) with the effective carrier temperature ($T_C$) and $\eta$ extracted with the proposed model.